\documentclass[aps,prl,twocolumn,superscriptaddress]{revtex4}

 \usepackage{amsmath,bm}
 \usepackage{mathrsfs}
 \usepackage{amsfonts}
 \usepackage{graphicx}
 \usepackage{ulem}
 \usepackage{setspace} 
 \usepackage{graphicx}
 \usepackage{epstopdf}
 \usepackage{dcolumn}
 \usepackage{amsmath}
 \usepackage{epsfig}
 \usepackage{indentfirst}
 \usepackage{psfrag}
 \usepackage{subfigure}
 \usepackage{amssymb}
 \usepackage{color}
 \usepackage{units} 
\usepackage{lineno}

 \usepackage{graphicx}
 \usepackage{dcolumn}
 \usepackage{bm}
 \usepackage{natbib}

\usepackage{physics}
\usepackage{dcolumn}
\usepackage{bm}
\usepackage{hyperref}
\usepackage{natbib}

\newcommand{\Figref}[1]{Fig.~\ref{#1}}
\def\b2{2B}

\usepackage{graphicx}
\usepackage[separate-uncertainty = true,multi-part-units=single]{siunitx}

\newcommand{\rev}[1]{{\color{black} #1}}
\makeatletter
\newcommand\colorsout[1]{\bgroup \markoverwith{\textcolor{#1}{\rule[0.5ex]{2pt}{0.4pt}}}\ULon}

\makeatother

\begin{document} 

\title{Magnetism of topological boundary states   induced by \\ boron substitution in graphene nanoribbons} 
 
    \author{Niklas Friedrich} \thanks{These two authors contributed equally to this work}
         \affiliation{CIC nanoGUNE, 20018 Donostia-San Sebasti\'an, Spain}
 
    \author{Pedro Brandimarte} \thanks{These two authors contributed equally to this work}
        \affiliation{Donostia International Physics Center (DIPC), 20018 Donostia-San Sebasti\'an, Spain}
  
    \author{Jingcheng Li}
        \affiliation{CIC nanoGUNE, 20018 Donostia-San Sebasti\'an, Spain}
  
  \author{Shohei Saito}
        \affiliation{Graduate School of Science, Kyoto University, Kyoto 606-8502, Japan}
  
  \author{Shigehiro Yamaguchi}
        \affiliation{Graduate School of Science, Nagoya University, Nagoya 464-8602, Japan}
  
  \author{Iago Pozo}
        \affiliation{CiQUS, Centro Singular de Investigaci\'on en Qu\'{\i}mica Biol\'oxica e Materiais Moleculares, 15705 Santiago de Compostela, Spain}
  
  \author{Diego Pe\~na}
        \affiliation{CiQUS, Centro Singular de Investigaci\'on en Qu\'{\i}mica Biol\'oxica e Materiais Moleculares, 15705 Santiago de Compostela, Spain}

  \author{Thomas Frederiksen}
        \affiliation{Donostia International Physics Center (DIPC), 20018 Donostia-San   Sebasti\'an, Spain}
        \affiliation{Ikerbasque, Basque Foundation for Science, Bilbao, Spain}
  
  \author{Aran Garcia-Lekue}
        \affiliation{Donostia International Physics Center (DIPC), 20018 Donostia-San Sebasti\'an, Spain}
        \affiliation{Ikerbasque, Basque Foundation for Science, Bilbao, Spain}
  
  \author{Daniel S\'anchez-Portal}\email{daniel.sanchez@ehu.eus}
        \affiliation{Donostia International Physics Center (DIPC), 20018 Donostia-San Sebasti\'an, Spain}
        \affiliation{Centro de F\'{\i}sica de Materiales CSIC-UPV/EHU, 20018 Donostia-San Sebasti\'an, Spain}

  \author{Jos\'e Ignacio Pascual} \email{ji.pascual@nanogune.eu}
        \affiliation{CIC nanoGUNE, 20018 Donostia-San Sebasti\'an, Spain}
        \affiliation{Ikerbasque, Basque Foundation for Science, Bilbao, Spain}

\begin{abstract}
Graphene nanoribbons (GNRs), low-dimensional platforms for carbon-based electronics, show the promising perspective to also incorporate spin polarization in their conjugated electron system. However, magnetism in GNRs is generally associated to localized states around zigzag edges, difficult to fabricate and with high reactivity. Here we demonstrate that magnetism can also be induced away from physical GNR zigzag edges through atomically precise engineering topological defects in its interior. A pair of substitutional boron atoms inserted in the carbon backbone breaks the conjugation of their topological bands and builds two spin-polarized boundary states around. The spin state was detected in electrical transport measurements through boron-substituted GNRs suspended between tip and sample of a scanning tunneling microscope. First-principle simulations find that boron pairs induce a spin 1, which is modified by tuning the spacing between pairs. Our results demonstrate a route to embed spin chains in GNRs, turning them basic elements of spintronic devices.
\end{abstract}
 
\date{\today}
\maketitle 
 
In spite of being a diamagnetic material, graphene can develop a special class of magnetism via the polarization of its $\pi$-electron cloud. Such $\pi$-paramagnetism is less localized than the more conventional $d$- or $f$-magnetism, and can interact over longer distances. Magnetic graphene nanostructures thus offer promising perspectives for \textit{a la carte} engineering of interacting spin systems with applications in quantum spintronics devices \cite{Awschalom2013,Slota2018b,Lombardi2019,Sharpe2019}. The vision of graphene $\pi$-paramagnetism has been recently boosted by the development of on-surface synthesis (OSS) as a versatile bottom-up route. In OSS, nanoscale graphene flakes with customized shape and composition are fabricated over a metal substrate through the steered reactions between designed organic precursors \cite{Cai2010,Corso2018}. Solid evidence of magnetism in flakes with zigzag edges has been revealed in scanning tunneling spectroscopy experiments \cite{Li2019,Mishra2020,LiPRL20,Lawrence20}.
 
 \begin{figure}[!t]
 	\includegraphics[width=0.99\columnwidth]{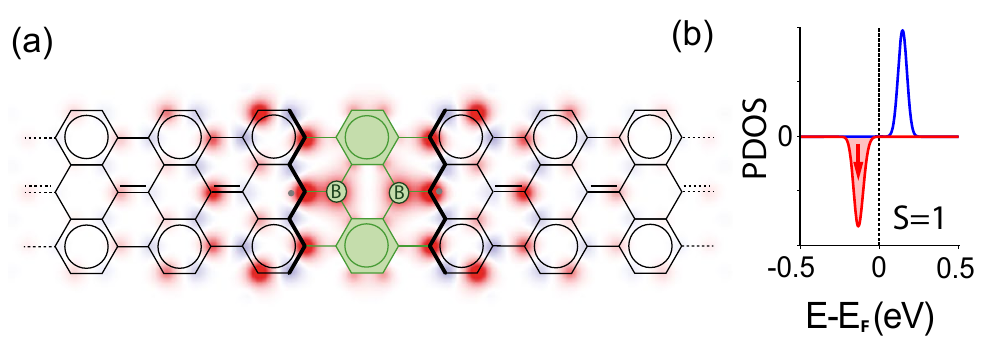}
 	\caption{(a) Lewis structure of the \b2-7AGNR  shown over a colour map representing  the spin polarization density map, computed by density functional theory simulations (\cite{SI}) (green   represents the boron moiety). (b) Spin-resolved Projected Density of States (PDOS) over carbon atoms around the boron dimer.  A net spin polarization of one kind confirms the ferromagnetic alignment of the two magnetic moments.}\label{fig:fig0}
 \end{figure} 
 
 Substituting one carbon atom of the graphene lattice by heteratoms is a potential route to induce magnetism\cite{Wang2014,Bonski2017}. 
 A representative case is the doping of graphene with substitutional boron atoms (\Figref{fig:fig0}), because it can be idealized as the removal of one electron from the conjugated bipartite lattice plus the energy upshift of a $p_z$ state. 
 However, boron atoms do not induce any spin imbalance around, but simply behave as a point potential \cite{Wang2014}. A prerequisite for the emergence of $\pi$-paramagnetism is that the point defect also causes a sufficiently large
 rupture of the conjugated electron system, for example by
 completely
 removing lattice sites or saturating $p_z$ orbitals \cite{Lieb1989,Palacios2008,Brihuega2016}, resulting in the localization of radical states.

 Here we show that inserting a pair of boron atoms in the carbon lattice of graphene nanoribbons (GNRs) enables a magnetic ground state. Density Functional Theory (DFT) simulations (\Figref{fig:fig0}) show that, while magnetism is completely absent around a pair of such substitutional B atoms in different sublattices of extended graphene, in a 7-carbon-wide armchair GNR (7AGNR) the boron pair builds up a net magnetic moment of 2$\mu_\text{B}$ (two Bohr magnetons). 
 The spin polarization,  shown in \Figref{fig:fig0}(a), decays towards the pristine segments with the characteristic shape of the 7AGNR end states \cite{Ijas2013} (see Supplementary Information (SI), \cite{SI} for a comparison).
 In fact, the spin cloud emerges from the rupture of the conjugated system imposed by the \b2-doped ring and the two neighbouring Clar sextets (green in \Figref{fig:fig0}(a)). 
 This moiety behaves as a highly reflective barrier for valence band electrons~\cite{Carbonell2017, Carbonell2018}, thus inducing localized end states associated with the termination of the topological 7AGNR valence band \cite{Cao2017}. 
 This striking result offers the vision of combining 
 band topology of nanoribbons~\cite{Cao2017,Rizzo2018,Groning2018} and heteroatoms for shaping spin textures in graphene ribbons.

 \begin{figure}[!t]
 	\includegraphics[width=0.8\columnwidth]{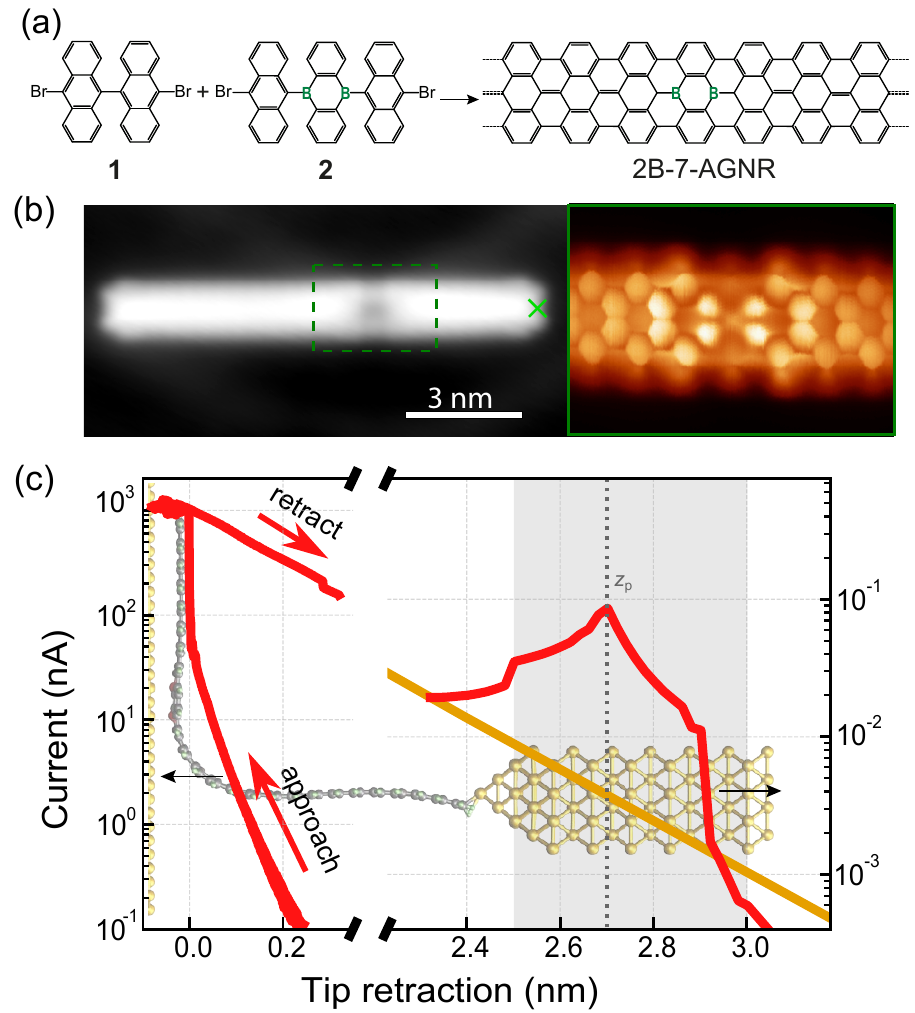}
 	\caption{ (a) Organic precursors mixed in the experiments.  
 		(b) STM constant current topography image of  a \b2-7AGNR ($V_\text{b} = \SI{-300}{mV}$, $I=\SI{30}{pA}$). The green cross indicates the position from where the GNR is lifted. (right) Constant height current scan ($V_\text{b} = \SI{2}{mV}$) using a CO-functionalized tip~\cite{Kichin2011} of the region indicated by the dashed rectangle.
 		(c) Tunneling current $I$ at $V_\text{b}=\SI{25}{mV}$ as a function of $z$ for a borylated (red) and a pristine (orange) GNR, for comparison. The grey region indicates where spectra in Fig.~3(b) was measured. The background shows results of atomistic simulations of a retraction stage  shown in Fig.~4, for illustration.}\label{fig:fig1a}
 \end{figure}

 In our experiments, we substitutionally inserted boron pairs (\b2) inside 7GNRs (\b2-7AGNRs) by adding a small fraction of 2B-doped trianthracene organic precursors (\textbf{1} in \Figref{fig:fig1a}(a))~\cite{Kawai2015,Cloke2015,Carbonell2017,Pedramrazi2018,Carbonell2018,Senkovskiy2018a} during the OSS of 7AGNRs using precursor \textbf{2}~\cite{Cai2010} (as schematically shown in \Figref{fig:fig1a}(a), see Methods in SI \cite{SI}). 
 Scanning tunneling microscopy (STM) images of the fabricated ribbons (\Figref{fig:fig1a}(b)) resolved the \b2 unit as a topography depression at varying positions inside the GNR~\cite{Kawai2015,Cloke2015,Carbonell2017}. Tunneling spectra showed no fingerprint of magnetism around the \b2 moieties due to the strong interaction between boron and metal states~\cite{Carbonell2018,Senkovskiy2018a}, which quenches the eventual magnetic ground state. 
 Therefore, to detect their intrinsic magnetic state the 2B moieties had to be removed from the metal substrate. 
  
\begin{figure}[!t]
	\includegraphics[width=0.8\columnwidth]{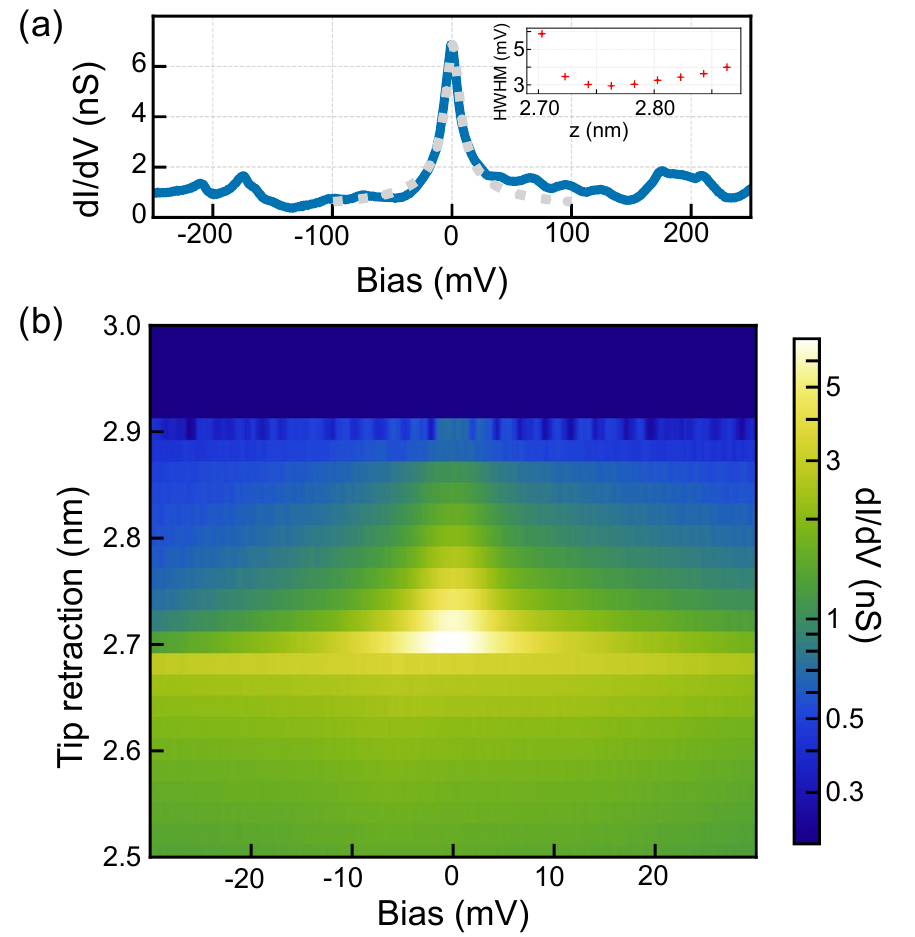}
	\caption{  (a)  Spectrum over a larger bias interval, taken at $z=\SI{2.70}{nm}$.
		The dotted grey line is a fitted Frota function \cite{Frota1992} with $\text{HWHM}=\SI{6\pm0.4}{mV}$. \rev{ The fitting interval is $|V_\text{b}|<\SI{20}{mV}$. The inset shows the evolution of the resonance's HWHM as a function of $z$.}
		(b) Conductance through the GNR as a function of $V_\text{b}$ and $z$.
		The corresponding height interval is the gray region indicated in Fig.2(c). A zero-bias, narrow resonance is observed for $\SI{2.7}{nm}<z<\SI{2.9}{nm}$.}\label{fig:fig1b}
\end{figure} 
 
 We used the STM tip to pick individual
 \b2-7AGNRs from one end (cross in \Figref{fig:fig1a}(b)) and lift them off to
 lie free-standing between tip and sample~\cite{Koch2012,Li2019}. 
 The (two-terminal) electrical transport through the suspended \b2-7AGNR was monitored during tip retraction $z$.  
 At the initial stages of suspension (\b2 unit still on the surface), the current  through the ribbon showed a weak exponential decrease with $z$ (\Figref{fig:fig1a}(c)), as for pristine GNRs~\cite{Koch2012}. However, at a certain retraction length $z_\text{p}$, the current exhibited a pronounced peak, returning afterwards to the previous exponential decay.
 The  peak and its  position $z_\text{p}$ were reproduced for several retraction/approach cycles of the same ribbon, and appeared in all \b2-7AGNRs studied. In every case, the value of $z_\text{p}$ correlated with the distance between the \b2 site and the contacted GNR-end (see SI~\cite{SI}),  proving that the current peaks were caused by the detachment of a \b2 moiety from the surface.

\begin{figure*}[!t]
	\includegraphics[width=0.9\textwidth]{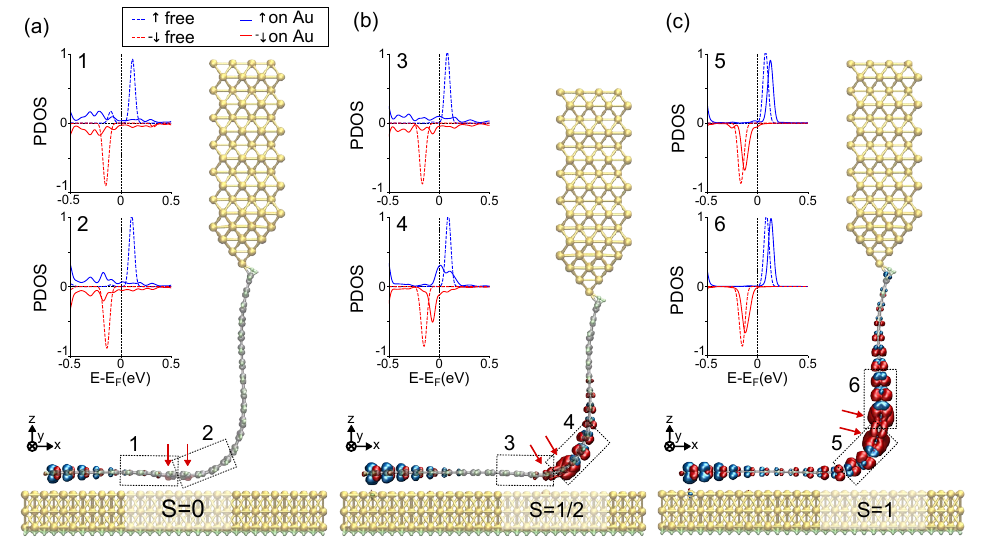}
	\caption{
		(a-c) Relaxed structures of three different configurations of a \b2-7AGNR bridging a gold tip and a Au(111) surface (red arrows indicate the position of the B heteroatoms).
		Constant spin density isosurfaces are shown over the atomic structure ($\SI{1.7e-3}{e/\angstrom^3}$, spin up in blue and down in red).
		Insets compare  spin PDOS over C atoms within the boxed regions around each boron atom for each bridge geometry (solid lines), with the corresponding one of a free \b2-7AGNR (dashed lines).
		The GNR zigzag termination on the surface holds a spin-polarized radical state, absent at the contacted end due the bond formed with the tip's apex \cite{Li2019}.
		\rev{ The equivalent PDOS are provided for a wider energy window in the SI.}
	}
	\label{fig:fig2}
\end{figure*}

 To explore the origin of the anomalous current peak, we measured differential conductance ($dI/dV$) spectra  at  positions around $z_\text{p}$ (\Figref{fig:fig1b}). The $dI/dV$) plots show the emergence of a narrow zero-bias  resonance in the spectra at $z_\text{p}=\SI{2.7}{nm}$, which gradually decreases its amplitude with tip retraction, and disappears for $z >\SI{2.9}{nm}$.
 The resonance remained pinned at zero bias in all the $z$ range observed. Its narrow line width reached a maximum value of $\Gamma_{\text{HWHM}}\approx \SI{6\pm0.4}{mV}$ at $z_\text{p}$, \rev{and evolved non-monotonously with retraction $z$} (\Figref{fig:fig1b}(a)) until disappearing. When the tip was approached below $z_\text{p}$, the resonance vanished abruptly, but it was recovered by increasing $z$ back above the $z_\text{p}=\SI{2.7}{nm}$ onset.
 From its narrow line shape and fixed zero-bias alignment, we conclude that the resonance is a manifestation of the Kondo effect \cite{kondo1964resistance,Temirov2008,Ternes2008spectroscopic}. 
 A Kondo-derived resonance appears in $dI/dV$ spectra when a spin polarized state weakly interacts with the conduction electrons of an underlying metal \cite{ternes2015spin}. 
 
 To correlate these observations  with the \b2-induced spin polarization predicted in   \Figref{fig:fig0}, we performed DFT simulations of a finite \b2-7AGNR suspended between a model gold tip and the surface of a Au(111) slab \cite{SI}. Figure \ref{fig:fig2} shows the relaxed atomic structures of the GNR-junctions before, while, and after detachment of a \b2 unit, and includes the computed constant spin density
 isosurfaces. Before \b2-detachment from the surface ($z<z_\text{p}$, \Figref{fig:fig2}(a)), the intrinsic magnetism around the \b2 units is quenched: the PDOS in the regions 1 and 2 around each boron atom is broad and spin unpolarized, contrasting with the clear spin polarization of free ribbons (shown as dashed plots). 
 This is caused by the strong hybridization of
 the B atoms with the gold surface~\cite{Kawai2015,Carbonell2018}, which appear $\SI{0.6}{\angstrom}$ closer to the surface than the carbon backbone.

 The detachment of the \b2 moieties from the metal surface causes the emergence of a net spin polarization, clearly reflected in their PDOS and spin density isosurfaces (\Figref{fig:fig2}(b,c)). At the intermediate snapshot of \Figref{fig:fig2}(b), only one of the two B heteroatoms is detached from the surface, and the ribbon hosts a net spin $S=1/2$ extending towards the free-standing segment (region 4). 
 For the fully detached \b2 case (\Figref{fig:fig2}(c)), both regions around the two B atoms (regions 5 and 6) are spin polarized, recovering the $S=1$ state of the isolated \b2-7AGNR (\Figref{fig:fig0}(b)). 
 \rev{Based on these simulations, we interpret that the most probable origin of the experimental Kondo resonance is the intermediate configuration pictured in \Figref{fig:fig2}(b).
 There, the Kondo effect is caused by the spin 1/2 of region 4 interacting weakly with the surface when the first boron atom is detached.
 Although the S=1 state of \Figref{fig:fig2}(c) could also produce a Kondo state \cite{LiPRL20}, one would expect that it shows a larger extension and is accompanied by inelastic triplet-singlet side bands. Instead, the zero-bias resonance in the experiments  disappears abruptly after a second kink $\sim$2~\AA\ higher (Figs. 2c and 3b) that we associate with the cleave of the second B atom, in consistency with the B-B distance.  The experimental results are thus consistent with the spin polarization around free-standing \b2 moieties. } 
 
 \begin{figure*}[!t]
 	\centering
 	\includegraphics[width=0.9\textwidth]{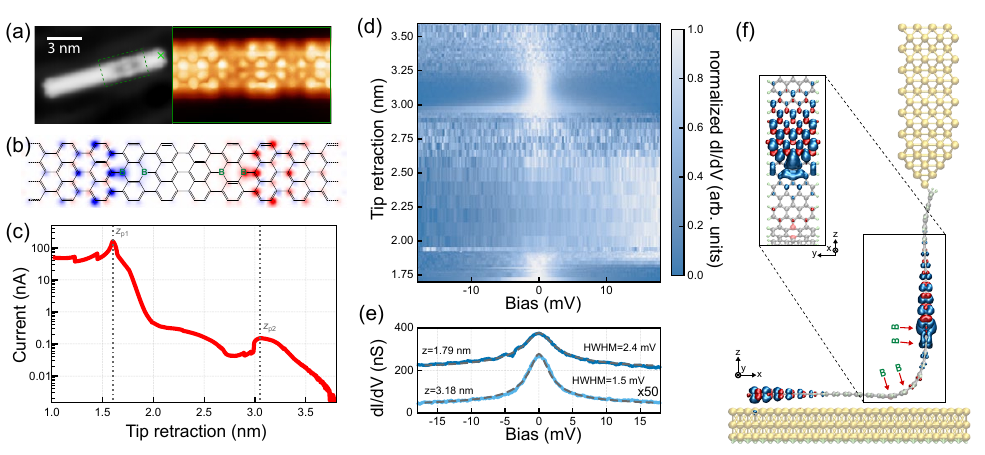}
 	\caption{ (a) (Left) Constant current STM image of a borylated GNR ($V_\text{b} = \SI{-300}{mV}$, $I=\SI{30}{pA}$).  (Right) Constant height current image of the marked rectangular region ($V_\text{b} = \SI{2}{mV}$) using a CO-functionalized tip.   (b) DFT simulation of the magnetization of a (\b2)$_2$-7AGNR. (c) Cotunneling current $I$ vs. $z$ through the ribbon in (a) suspended between tip and sample.  (d) Normalized differential conductance of the suspended (\b2)$_2$-7AGNR as a function of $V_\text{b}$ and $z$ (see SI~\cite{SI}). Two zero-bias resonances are observed. (e) Representative $dI/dV$ spectra measured at the indicated $z$ positions, with  fits (dashed) using Frota functions~\cite{Frota1992}. (f) DFT relaxed structure of a suspended (\b2)$_2$-7AGNR. Constant spin-density isosurfaces are shown over the atomic structure; red arrows indicate the position of the B atoms. The inset shows the indicated region of the suspended GNR from a different angle.}\label{fig:fig3}
 \end{figure*}

 Although the Kondo signal vanishes quickly with retraction, DFT finds that the $S=1$ state of the free ribbon remains, and is clearly favored over an anti-parallel alignment by $\sim \SI{14}{meV}$ per isolated \b2 pair. 
 The presence of a triplet state is striking; the two spin clouds at each side of the \b2 center extend symmetrically over opposite sub-lattices of the 7AGNR,  what usually favors an antiparallel kinetic exchange~\cite{Brihuega2016}. A detailed analysis reveals that the hopping matrix elements between the two localized states at the sides of one \b2 unit are very small ($t_\text{intra}\sim 18$~meV, see SI~\cite{SI}). Consequently, the moiety formed by a \b2-doped ring surrounded by two Clar sextets is a very stable element that blocks conjugated electrons from hopping across. This explains the presence of a magnetic state because the borylated element acts as a barrier for valence band electrons of the 7AGNR segments~\cite{Carbonell2017}, and induces spin polarized boundary states due to the non-trivial topology of this band~\cite{Cao2017}.
 Additionally, the \b2 barrier also disconnects the boundary states at each side, and hinders the (anti-parallel) kinetic exchange between them.
 The stabilization of the triplet configuration is then the result of the weak direct overlap 
 between both spin-polarized boundary states through the \b2 barrier, which, due to the tiny hopping between them, dominates the exchange interaction and induces the ferromagnetic alignment of the spins according to Hund's rule. 
 
 We also studied GNRs with two consecutive \b2 moieties like  in \Figref{fig:fig3}(a,b), spaced by $\SI{1.2}{nm}$. Transport experiments through these GNRs  as a function of tip-sample distance (\Figref{fig:fig3}(c))  also reveal  deviations from an  exponential decay with $z$, but now showing two peak features at retraction distances $z_\text{p1} \approx \SI{1.60}{nm}$ and $z_\text{p2} \approx \SI{3.05}{nm}$. These values are related to the positions of the \b2 units (nominally $\sim 2.0$ and  $\sim\SI{3.2}{nm}$ from the contact point, respectively).
 A map of (normalized) differential conductance as a function of bias and $z$ (\Figref{fig:fig3}(d)) shows that both current features are also caused by narrow zero-bias $dI/dV$ resonances appearing at ranges $z<\SI{1.9}{nm}$ and $\SI{2.9}{nm} < z < \SI{3.5}{nm}$, respectively. Their line shape (\Figref{fig:fig3}(e)) is similar to the resonances observed for the single \b2 case, and can also be attributed to Kondo states, which reflect the emergence of spin-polarization in the ribbon as each \b2 unit is detached from the surface.

 Although these results apparently suggest that each \b2  behaves as an independent spin center, DFT simulations of free (\b2)$_2$-7AGNRs (\Figref{fig:fig3}(b)) find that 
 the two singly-occupied boundary states  between 
 neighboring \b2 elements interact strongly and open a large hybridization gap~\cite{Carbonell2018}, forcing them into a closed shell configuration. As a consequence, 
 the spin polarization vanishes  between two \b2 sites, but persists outside this region as two uncompensated spin 1/2 clouds (\Figref{fig:fig3}(b)) with barely no preferred relative spin alignment. 
 From our electronic structure calculations~\cite{SI,Carbonell2018} we can characterize this hybridization by a relatively large effective hopping term $t_\text{inter}$ between boundary states of neighboring \b2 units, which contrasts with the weak hopping $t_\text{intra}$ across each \b2 unit.
 In fact,  the electronic structure close to the Fermi level of a sequence of borylated units can be mapped  onto the Su-Schrieffer-Heeger (SSH) model~\cite{SSH1979}, characterized by two alternating hoppings along a 1D wire. Since $t_\text{inter} > t_\text{intra}$, an alternative way to understand the spin-polarized states in \Figref{fig:fig3}(b) is as zero-energy topological modes of a very short SSH chain. These simulations allows us to predict that a S=1 spin chain will emerge for larger inter-\b2 spacing, when both hopping terms become smaller than the Coulomb charging energy $U$ of the boundary states~\cite{SI}.  
 
 \rev{To explore if inter-2B interactions survive in the experimental geometry, we simulated a (\b2)$_2$-7AGNR suspended between tip and sample. \Figref{fig:fig3}(f) show the spin polarization of a snapshot with one \b2 moiety  completely detached and the second partially bound to the surface. In contrast with the large spin cloud around the lifted single \b2 in \Figref{fig:fig2}(c), here there is no spin density between the two \b2 moieties, but a net S=1/2 cloud above, confirming the presence of interactions between 2B units.  
 The second S=1/2 boundary cloud  is expected to appear below the lower \b2 only after the last boron atom is detached, being this responsible for the more extended Kondo effect observed in the experiment above $z_\text{p2}$.} These results confirm the spin polarization predicted at the interface between (\b2)$_2$ units and pristine 7AGNR segments~\cite{Cao2017}, which in essence are zero-energy modes of the 7AGNR valence band of similar nature than those created by a single \b2 unit at every side.
 
 The peculiar spin polarization of single \b2 units and dimers is a remarkable consequence of the large and long-range exchange interactions present in GNRs~\cite{Cao2017,Li2019,Mishra2020}. Our results support that  spins survive in free-standing GNRs and can form $S=1$ spin chains for low concentrations. Tuning the separation between \b2 units is a promising strategy to control spin polarization through a change in the correspondence between inter-\b2 and intra-\b2 interactions. Furthermore, the stronger sensitivity of substitutional boron heteroatoms to chemical bonding endows these systems with ideal properties to manipulate complex spin states in chains.

\begin{acknowledgments}
	We gratefully acknowledge financial support from Spanish AEI (MAT2016-78293-C6, FIS2017-83780-P, and the Maria de Maeztu Units of Excellence Programme MDM-2016-0618), from the European Union (EU) through Horizon 2020 (FET-Open project SPRING Grant. no. 863098), the Basque Departamento de Educaci\'on through the PhD fellowship No.~PRE$\_$2019$\_$2$\_$0218 (S.S.), the Xunta de Galicia (Centro singular de investigaci\'on de Galicia 
	accreditation 2016-2019, ED431G/09), the University of the Basque Country (Grant IT1246-19), and the European Regional Development Fund (ERDF). I.P. also thanks Xunta de Galicia and European Union (European Social Fund, ESF) for the award of a predoctoral fellowship.
\end{acknowledgments}

\bibliographystyle{apsrev4-1}  

%

\end{document}